\documentclass[twocolumn,showpacs,aps,prl]{revtex4}

\usepackage{graphicx}%
\usepackage{dcolumn}
\usepackage{amsmath}
\usepackage{latexsym}

\begin {document}

\title
{
Branching Process in a Stochastic Extremal Model
}
\author
{
S. S. Manna
}
\affiliation
{
\begin {tabular}{c}
Max-Planck-Institute f\"ur Physik Komplexer Systeme,
    N\"othnitzer Str. 38, D-01187 Dresden, Germany \\
Satyendra Nath Bose National Centre for Basic Sciences
    Block-JD, Sector-III, Salt Lake, Kolkata-700098, India
\end{tabular}
}
\begin{abstract}

      We considered a stochastic version of the Bak-Sneppen model (SBSM) of ecological evolution where the
   the number $M$ of sites mutated in a mutation event is restricted to only two. Here the mutation zone consists 
   of only one site and this site is randomly selected from the neighboring sites at every mutation event in 
   an annealed fashion. The critical behavior of the SBSM is found to be the same as the BS model in 
   dimensions $d$ =1 and 2. However on the scale-free graphs the critical fitness value is non-zero 
   even in the thermodynamic limit but the critical behavior is mean-field like. Finally $\langle M \rangle$
   has been made even smaller than two by probabilistically updating the
   mutation zone which also shows the original BS model behavior. We conjecture that a SBSM on any arbitrary 
   graph with any small branching factor greater than unity will lead to a self-organized critical state.

\end{abstract}
\maketitle

      In a seminal paper Bak and Sneppen introduced a Self-Organized Critical (SOC) \cite {Bak} 
   model for the ecological evolution of interacting species, known as the Bak-Sneppen (BS) 
   \cite {Bak-Sneppen} model. In this model an entire species is represented by a single fitness
   variable. Using the spirit of Darwinian principle the minimally fit species is mutated. This however
   disturbs the stability of the ecological system. There are some other species which are dependent 
   on the minimally fit species for example as a part of the food web. These species are also mutated.
   The ecological evolution takes place in a series of such events.

      The phenomenon of SOC is the spontaneous emergence of fluctuations of all length and time
   scales in a slowly driven system. This concept was first introduced to describe the
   formation of a sandpile of a fixed shape \cite {Bak}. Later the idea of SOC has been applied 
   to a large number of different physical systems \cite {Bakbook}. A number of models have been
   introduced to describe SOC in different systems. In the Bak, Tang and Wiesenfeld (BTW) model 
   \cite {Bak} the dynamics is described in terms of spreading of sand grains on a sandpile. 
   Toppling of an unstable sand column distributes sand grains to all neighboring sites. This 
   model is also known as the Abelian Sandpile Model since the stationary state is independent 
   of the sequence of grain additions \cite {Dhar}. In a stochastic version of the sandpile model 
   grains are distributed to randomly selected neighboring sites \cite {MannaSOC}. In the SOC models 
   fluctuations are described in terms of avalanches of activities and their size distributions 
   assume power law decaying functions for large system sizes. The BS model is regarded as a 
   simple but non-trivial SOC system.
 
      The BS model is described as follows. The ecosystem consists of $N$ species located at the 
   sites $i=1,N$ of an one dimensional lattice. A fitness variable $f_i$ is associated with every 
   site. Initially uniformly distributed random numbers within the range $\{0,1\}$ are assigned for 
   the fitness values. The dynamical evolution of the ecosystem takes place in a series of mutation 
   events. Each event consists of two steps: (i) The `active' site $i_o$ is searched out which has 
   the minimal fitness $f_o$. This site is mutated i.e., the value of $f_o$ is replaced by a new 
   random number. (ii) All sites of a fixed mutation zone in the local neighborhood are mutated 
   as well. E.g., in $d$=1 the fitness values at two neighboring sites of $i_o$ are also refreshed.
   These two steps complete a single mutation event. After that the active site is located at some 
   other site where the next mutation event takes place and so on. The sequential time is measured by the number 
   of mutation events. The system eventually reaches a steady state in which the associated statistical 
   distributions assume their time independent stationary forms. The recurrent culling of the globally 
   minimal fitness values leads to a step like form of the probability distribution $P(f)$ so that 
   in the limit of $N \to \infty$: $P(f)=0$ for $f < f_c$ and $P(f) = {\cal C}$ a constant otherwise, 
   where $f_c$ is a critical fitness threshold \cite {Video}. 

      In the steady state the fluctuations are described in terms of avalanches. A critical avalanche 
   is a sequence of successive mutation events with $f_o < f_c$. The life-time $s$ of the avalanche is 
   the total number of events in the avalanche. The distribution of the avalanche life-times has a 
   power law tail in the limit of $N \to \infty$ : $D(s) \sim s^{-\tau}$. The BS model has been studied 
   on hypercubic lattices, e.g., the values of $f_c$ and $\tau$ are found to be 0.66702(8) and 
   1.073(3) \cite {Grassberger} and 0.328855(4) and 1.245(10) \cite {Paczuski1,Dorogov} in $d$ 
   = 1 and 2 respectively. The upper critical dimension has been argued to be 4 \cite {Paczuski2} and 8 
   \cite {Rios} where $\tau$ assumes its mean-field value of 3/2.

      It has been observed that increasing the size of the mutation zone modifies the critical fitness 
   $f_c$ but not the critical behavior \cite {Garcia}. It has also been shown that the BS model with isotropic 
   and the anisotropic mutation zones have different critical behaviors \cite {Maslov}. Variants of the 
   BS model with exponentially and power-law distributed random numbers have been studied \cite {Vergeles,Cafiero}. 
   BS model has also been studied on different heterogeneous graphs as well, e.g., on random graphs 
   \cite {Christensen} and on an adoptive networks \cite {Caldarelli}. However, the critical fitness 
   threshold is zero for BS model on infinitely large scale-free graphs \cite {Moreno}. In a 
   scale-free graph the degree 
   distribution $P(k)$ decays as a power law (degree $k$ being the number of edges meeting a vertex) as: 
   $P(k) \sim k^{-\gamma}$ and the cut-off $k_{max} \sim N^x$, $N$ being the size of the graph. The 
   Barab\'asi-Albert (BA) network \cite {Barabasi} is a well-known scale-free graph with $\gamma=3$ and 
   $x=1/2$ \cite {BarabasiRMP}. For the BS model on BA network, the value of $f_c(N)$ decreases to zero 
   in $N \to \infty$ limit as $1/\log(N)$ \cite {Moreno, Lee, Lee1, Masuda}.

\begin{figure}[top]
\begin{center}
\includegraphics[width=6.5cm]{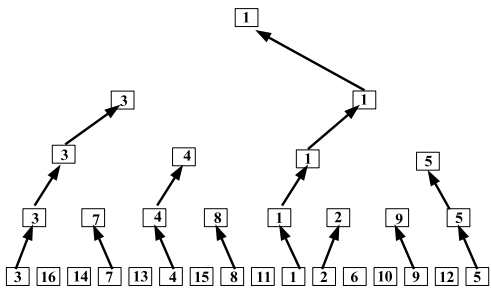}
\end{center}
\caption{
Hierarchical organization of the data structure to search for the global minimum 
\cite {Grassberger}. An $1d$ array of 16 sites has 16 random numbers. Successive 
pairs of sites form blocks in level 1. The smaller number of every block is 
forwarded to the level 2 and a pointer is attached. Similar block pairs are also 
formed in level 2 and pointers are attached towards level 3 and so on. One moves 
opposite to the pointer directions starting from the top level to reach the minimal 
site at the lowest level.
}
\end{figure}

      First, let us consider two limiting cases. Suppose the mutation zone has size zero
   so that only the active site is updated in every mutation event. Then the chance that the
   next $f_o$ will be less than the present $f_o$ arises due to refreshing this site only. 
   Consequently $f_o$ increases almost monotonically leading to $f_c=1$. The other limiting 
   case is the $N$-clique graph where each vertex is connected to all other $N-1$ vertices 
   in the graph \cite {N-clique}. The mutation zone consists of $N-1$ nodes and therefore in 
   a mutation event fitness values of all $N$ vertices are refreshed, as a result both 
   $f_o, f_c \to 0$ as $N \to \infty$. Therefore when the size of the mutation zone is in 
   between 0 and $N-1$, there is a competition between the mechanisms of these two limiting 
   processes and consequently $f_c$ assumes a non-trivial value between $\{0,1\}$.

      Since a random graph has Possionian degree distribution which decays very fast the 
   threshold fitness $f_c$ has a fixed value \cite {Christensen}. On the other 
   hand in scale-free networks whenever a mutation event initiates at a hub vertex its large 
   number $\sim N^x$ of neighbors are refreshed. Qualitatively this is a similar mechanism as 
   the $N$-clique graph, but since $x < 1$, the $f_c(N) \to 0$ inverse logarithmically.  
   Therefore if the size of the mutation zone is reduced, number of refreshed sites are less, 
   consequently the chance of creating new $f_o$ smaller than the present minimal is less. As a 
   result $f_c$ goes up.

\begin{figure}[top]
\begin{center}
\includegraphics[width=6.5cm]{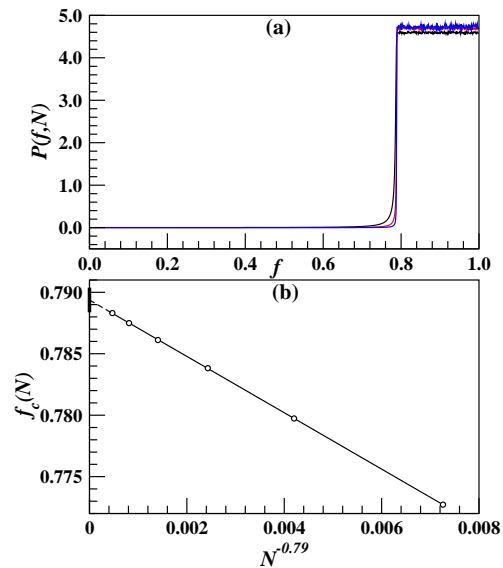}
\end{center}
\caption{(Color online)
(a) The distribution $P(f,N)$ of fitness values are shown for $1d$ SBSM for system sizes 
$N = 2^{10}$ (black), $2^{12}$ (red) and $2^{14}$ (blue). The jump in the distribution 
becomes gradually sharper with increasing the system sizes.
(b) The threshold fitness values for different system sizes are plotted with $N^{-0.79}$ 
and extrapolated to obtain the $f_c(\infty)$= 0.7894(10). The solid line is a least square 
fit of the data and the thick line indicates the extent of the error.
}
\end{figure}

      The dynamics in the Bak-Sneppen model is well known to be described by a branching process 
   \cite {Paczuski1,Felici}. A branching process \cite {Kim} is defined by a population where 
   each individual in one generation produces randomly a number of offsprings in the next generation. 
   The average number of offsprings is called the basic reproduction rate. Here in BS model we call 
   the sites with fitness values $f < f_c$ as the critical sites. Every mutation event produces 
   randomly a number of critical sites. If the total number of sites refreshed in a mutation event 
   is $M$, which is the number of sites in the mutation zone plus one for the active site, then the 
   average number of critical sites produced in a mutation event is $r_b = Mf_c$ which we call as 
   the branching factor. It may be noted that all critical sites produced in a mutation event may 
   not be fresh critical sites. An existing critical site may be refreshed again to a critical site 
   with a different fitness value. It is known that a non-trivial branching process needs a value 
   of the basic reproduction rate greater than unity. For BS model a similar condition may be that 
   the branching factor should be greater than one. This implies that if one reduces the size of the 
   mutation zone to only one site, it may still be possible to achieve a Self-organized Critical 
   state only if $r_b = 2f_c > 1$. In the following we will present numerical evidence which indicates that indeed 
   this is likely to be true. In a particular anisotropic case the fixed one member mutation zone 
   has already been studied in one dimension \cite {Garcia}.

      In this paper we study a stochastic Bak-Sneppen model (SBSM) with the simplest possible mutation zone. 
   In addition to the usual procedure of randomly refreshing the fitness values the stochasticity is introduced 
   in randomly selecting the sites of the mutation zone as well. The size (the number of sites) of the mutation zone 
   is kept fixed at the minimal value i.e., unity and for different mutation events different mutation zones are 
   randomly selected only from the nearest neighbors of the active site.
      
\begin{figure}[top]
\begin{center}
\includegraphics[width=6.5cm]{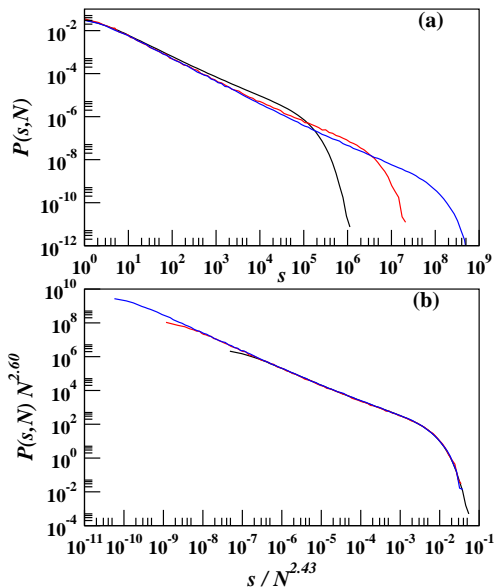}
\end{center}
\caption{(Color online)
The avalanche life-time distributions $P(s,N)$ for the $1d$ SBSM. In (a) we show the plots for the 
system sizes $N$ = $2^{10}$ (black), $2^{12}$ (red) and $2^{14}$ (blue) (from left to right).
(b) A finite-size scaling of this data shows an excellent data collapse for $\eta_1$ = 2.60 and 
$\zeta_1$ = 2.43 giving the value of the avalanche life-time exponent to be $\tau_1 \approx 1.07$. 
The left end of each curve shifts to the left on increasing system size.
}
\end{figure}

      We follow the Grassberger algorithm for our study \cite {Grassberger}. Searching for the minimal 
   fitness using a brute-force algorithm needs to test the fitness values of all sites requiring CPU $\sim N$. 
   Grassberger used a hierarchical organization of the data in block structure where CPU $\sim log(N)$. 
   Sites of an one dimensional lattice of $N=2^n$ sites are divided into $N/2$ block pairs like (1,2), 
   (3,4),... ($N-1,N$). For each block, the smaller fitness value of two sites is stored in a site of 
   another lattice of $N/2$ sites in a higher level and a pointer is assigned to this site. This procedure 
   is repeated for the next higher level as well. Finally only one site in the $(n+1)$-th level contains 
   the global minimal fitness value (Fig. 1). To locate the minimal fitness site one moves opposite to 
   the pointer directions starting from the top level. During the mutation of every site at the lowest 
   level one needs to update the fitness values and the pointer directions upto the top level. 

      Indeed we observe that the simple SBSM preserves all characteristics of the
   original BS model. In $d$=1, the step form of $P(f)$ has been observed in all system
   sizes from $N = 2^7$ to $2^{14}$, increased by a factor of 2. The jump in $P(f)$ at $f_c(N)$ is rather 
   smooth for the small $N$ but becomes more and more sharper on increasing $N$ (Fig. 2(a)). The 
   $f$ value at the intersection of two successive $P(f)$ curves gradually shifts to higher values with $N$. 
   The value of $f_c(\infty)$ is estimated as follows. In $N \to \infty$ limit the 
   normalization of $P(f)$ gives ${\cal C}(\infty) = 1/(1-f_c(\infty))$. The average fitness per site is 
   then $\langle f(\infty) \rangle = \int^1_0 fP(f)df = ({\cal C}(\infty)/2)(1-f^2_c(\infty)) = (1+f_c(\infty))/2$
   which gives $f_c(\infty) = 2\langle f(\infty) \rangle -1$.
   The whole fitness profile is sampled at the interval of every $N$ mutation events
   and $f_c(N)$ values are extrapolated with $N^{-\kappa}$ with $\kappa = 0.79 \pm 0.01$ to obtain 
   $f_c(\infty)=0.7894(10)$ (Fig. 2(b)). The most suitable value of $\kappa$ is decided 
   by trying different trial values of it and then selecting that particular value for which the fitting error is minimum.
   On a similar extrapolation of $\langle f(N) \rangle$s we get $\langle f(\infty) \rangle = 0.8947(10)$ implying that the
   branching factor $r_b = 1.5788 > 1$.

\begin{figure}[top]
\begin{center}
\includegraphics[width=6.5cm]{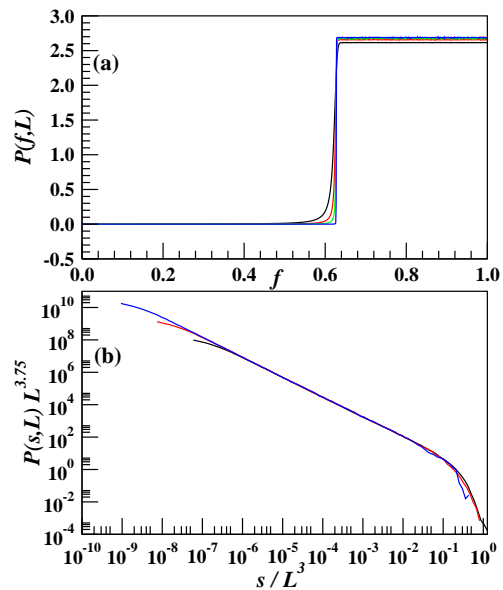}
\end{center}
\caption{(Color online)
Data for the $2d$ square lattice of size $L$. (a) The fitness distributions $P(f,L)$ vs. $f$ for the 
system sizes $L$ = $2^5$ (black), $2^6$ (red), $2^7$ (green) and $2^{10}$ (blue). The jump in the 
distribution becomes gradually sharper with increasing the system sizes.
(b) The finite size scaling of the avalanche size distribution for $L$ = $2^8$ (black), $2^9$ (red) 
and $2^{10}$ (blue) giving $\eta_2=3.75$, $\zeta_2=3.00$ and $\tau_2=1.25$. The left end of each 
curve shifts to the left on increasing system size.
}
\end{figure}

      The life-time distribution exponent $\tau$ is estimated using the method of finite-size
   scaling analysis. Large number of mutation events are studied: $\approx 4\times10^{10}$ for $N=2^7$
   to  $\approx 110\times10^{10}$ for $N=2^{14}$. In the Fig. 3(a) the binned probability $P(s,N)$ 
   distribution data for $N = 2^{10}, 2^{12}$ and $2^{14}$ are only plotted. A direct measurement
   of the slopes of these three curves gives 0.94, 1.02 and 1.05 for the estimates of $\tau$ for the 
   three system sizes respectively. A finite-size scaling analysis has been done using the following 
   scaling form:
\begin {equation}
P(s,N) \propto N^{-\eta} {\cal G}(s/N^{\zeta})
\end {equation}
   where the scaling function ${\cal G}(x) \sim x^{-\tau} $ in the limit of $x \to 0$ and 
   ${\cal G}(x)$ approaches zero very fast for $x >> 1$. The exponents $\eta$ and $\zeta$ fully 
   characterize the scaling of ${P(s,N)}$ in this case. An immediate way to check the validity 
   of this equation is to attempt a data collapse by plotting $P(s,N)N^{\eta}$ vs. $s/N^{\zeta}$ 
   with trial values of the scaling exponents. The values for obtaining the best data collapse are found 
   to be $\eta_1=2.60$ and $\zeta_1=2.43$, here we have used the subscripts to denote the 
   dimension of the system (Fig. 3(b)). The life-time exponent for $1d$ SBSM is therefore $\tau_1=\eta_1/\zeta_1 
   \approx 1.07(2)$. This exponent is very close to the value of $\tau_1$ in the BS 
   model \cite {Grassberger}.
   
\begin{figure}[top]
\begin{center}
\includegraphics[width=6.5cm]{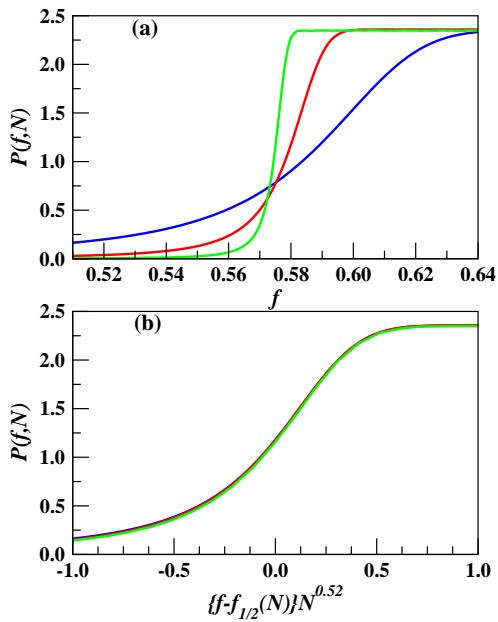}
\end{center}
\caption{(Color online)
The distributions $P(f,N)$ of the fitness values in the steady state of SBSM on the Barab\'asi-Albert 
scale-free graph. (a) Distribution plots for the graph sizes $N = 2^7$ (blue), $2^{10}$ (red) and 
$2^{13}$ (green). The jump in the distribution becomes gradually sharper with increasing the system sizes.
(b) A finite-size scaling of this data shows an excellent collapse.
}
\end{figure}

      Next, we have studied the SBSM on a two dimensional square lattice of size $L \times L$, so that
   total number of sites $N = L^2$. The fitness distribution profiles have been shown in Fig. 4(a) for 
   different system sizes. The critical fitness value $f_c$ has been obtained as 0.628(1) on extrapolating 
   the $f_c(L)$ values of $L=2^7, 2^8, 2^9$ and $2^{10}$ with $L^{-1.29}$ and $r_b = 1.256 > 1$. Another 
   finite-size scaling has been done in a similar way to obtain the scaling function exponents for the 
   life-time distribution as shown in Fig. 4(b). The values obtained for the best data collapse are 
   $\eta_2=3.75$ and $\zeta_2=3.0$ yielding the value of $\tau_2=1.25(2)$. This exponent is also very 
   close to the value of $\tau$ obtained for two dimensional BS model \cite {Paczuski1,Dorogov}.

      The SBSM is also studied on the scale-free BA network. In this growing graph every new vertex comes up
   with $m$ edges and gets connections to $m$ distinct vertices of the existing graph. Initially the
   growth starts from a $(m+1)$ clique. The actual growth process is executed by the improved
   algorithm \cite {Manna}. In this method a new vertex selects one of the existing edges with uniform 
   probability and gets a connection to one of its end vertices with probability 1/2 to generate the
   BA graph. For every vertex this process is repeated $m$ times for attaching $m$ links keeping track
   that all $m$ vertices must be distinct. We have used $m = 2$ in our studies.

\begin{figure}[top]
\begin{center}
\includegraphics[width=6.5cm]{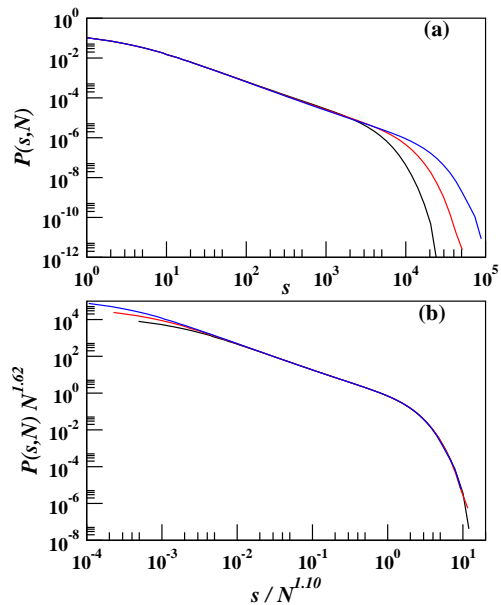}
\end{center}
\caption{(Color online)
The avalanche life-time distributions $P(s,N)$ of the SBSM on the BA graphs. In (a) we show the 
plots for the system sizes $N = 2^{10}$ (black), $2^{11}$ (red) and $2^{12}$ (blue) (from left to
the right).
(b) A finite-size scaling of this data shows an excellent data collapse for $\zeta_{BA}$ = 1.62 and 
$\eta_{BA}$ = 1.10 giving the value of the avalanche life-time exponent to be $\tau_{BA} \approx 1.47$.
The left end of each curve shifts to the left on increasing system size.
}
\end{figure}

      For the SBSM on a BA 
   graph, the mutation zone is a single randomly selected vertex out of all the $k$ neighboring vertices.
   As expected this system also gradually evolves to a steady state. The stationary fitness distribution $P(f)$
   shows up the characteristic jump at a certain value of $f_c$. For the small graphs $P(f)$ grows
   continuously across the critical fitness value from a low value to a high value.
   All quantities measured are averaged over different independent realizations of BA graphs.
   Both the fitness function and avalanche size distribution differ little from one graph to
   the other so that when the number of configurations are increased the fluctuations in the averaged plot gradually
   reduced.
   The dynamics is followed till $\approx 12 \times 10^{10}$ mutation events including
   2500 ($N=2^7$) to 152 ($N=2^{13}$) un-correlated BA graphs. The arrival of the steady 
   state is ensured
   by keeping track of the average fitness value $\langle f \rangle$ per vertex which initially
   grows but eventually becomes steady. The stationary state data is collected skipping the first 25 
   million mutation events as the relaxation time. For the fitness distribution, the fitness data is 
   collected from all vertices of the network at the interval of every $N$ mutation events. In Fig. 5(a)
   the fitness distribution $P(f)$ has been plotted for the system sizes $N = 2^7, 2^{10}$ and $2^{13}$.
   The average fitness values $\langle f(N) \rangle$ are measured to obtain the critical fitness
   thresholds for different $N$ values. These on extrapolation with $N^{-4/3}$ gives $f_c = 0.5751(10)$.
   Therefore for SBSM on BA graph the branching factor $r_b = 1.1502 > 1$.

\begin{figure}[top]
\begin{center}
\includegraphics[width=6.5cm]{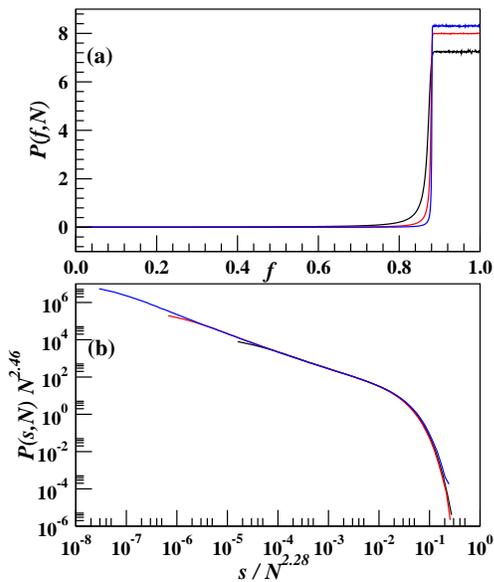}
\end{center}
\caption{(Color online)
Data for the case when the average number of sites updated in a mutation event $\langle M \rangle$ = 1.5 in $1d$
for the system sizes $N = 2^7$ (black), $2^9$ (red) and $2^{11}$ (blue). 
(a) The fitness distributions $P(f,N)$ vs. $f$.
The
jump in the distribution becomes gradually sharper with increasing the system sizes.
(b) The finite size scaling of the avalanche size distribution for the same system sizes
giving $\eta=2.46$, $\zeta=2.28$ and $\tau \approx 1.08$.
The left end of each curve shifts to the left
on increasing system size.
}
\end{figure}

      For BA graph we have done a finite-size scaling of the fitness distribution data to exhibit explicitly that
   the value of the fitness threshold $f_c(\infty)$ is indeed non-zero even in the thermodynamic limit.
   To make a data collapse we need to stretch different curves of Fig. 5(a) to different amounts along the
   $f$ axis. For that we need to keep one point fixed on every curve. We selected this point which has 
   $P(f,N) = {\cal C}(N)/2$. The corresponding value of $f$ is denoted by $f_{1/2}(N)$ and is calculated by 
   an interpolation. The $f$ axes are then shifted by $f_{1/2}(N)$ amounts for all three plots and
   have been rescaled by $N^{0.52}$ to obtain a nice data collapse which is shown in Fig. 5(b).

      The avalanche size distributions have been studied for the same values of $f_c(N)$ obtained 
   above. The raw distribution data has been shown in Fig. 6(a) for the three graph sizes $N = 2^{10}, 2^{11}$ and $2^{12}$. 
   A similar finite-size scaling analysis yields $\eta_{BA}$ = 1.62 and $\zeta_{BA}$ = 1.10
   giving the value of the life-time exponent $\tau_{BA} \approx 1.47(3)$ (Fig. 6(b)). This value is close to
   the mean-field value of 1.5 obtained by \cite {Moreno}.
   
      The branching factor can be reduced even further. During every mutation event we first 
   decide with a probability $p=1/2$ if we update the single site in the mutation zone or not. If it is
   favored only then one of the nearest neighbors is selected randomly and updated as in SBSM, otherwise
   the active site is only updated and therefore $\langle M \rangle$ = 1.5. In $d$=1, we get an enhanced 
   value of $f_c \approx 0.883(5)$ which means $r_b = \langle M \rangle f_c \approx 1.325 > 1$. The critical exponents 
   found to be very consistent with SBSM at $d$=1. Fig. 7(a) shows the fitness distribution plot whereas
   Fig. 7(b) exhibits the finite-size scaling analysis of the avalanche size distributions.
   Our expectation is that on further reduction of the branching factor by reducing $p$, the $f_c$ will 
   go up but the critical behavior would remain intact. To check it we simulated only one system size of 
   $N=2^{11}$ for $\langle M \rangle$ = 1.75, 1.25, 1.125 and 1.0625 and obtained $f_c(2^{11})$ values 
   0.828, 0.935, 0.966 and 0.982 respectively corresponding to branching factors 1.449, 1.168, 1.087 and 1.044
   respectively, all values larger than unity. Direct measurement of $\tau$ exponents from $D(s)$ vs. $s$ 
   plots gives 1.07, 1.05, 1.04 and 1.01 respectively. Therefore it seems likely that for any branching 
   factor greater than unity, the SBSM would exhibit a non-trivial SOC state.

      Lastly we studied the SBSM on the $N$-clique graph. Our single member mutation zone is a special case 
   of the random BS model studied in \cite {Flyvbjerg}. For all finite size systems the $f_c(N)$ values are 
   larger than 1/2 but approach to it as $N$ increases. The results in the asymptotic limit are consistent 
   with \cite {Flyvbjerg} i.e., $f_c=0.5$, $\eta_{NC}=1.5$ and $\zeta_{NC}=1.0$ giving $\tau_{NC} = 1.5$, a 
   complete mean-field behavior. Here $r_b = 1$ holds good only in the asymptotic limit, for all finite system
   sizes $r_b > 1$.

      To summarise, the ecological evolution process described in the Bak-Sneppen SOC model has been widely
   regarded as a branching process. Here each mutation event generates randomly on the average $r_b$ offsprings
   with fitness values under the threshold. Similar to the basic reproduction rate in the theory of branching 
   process we propose that a non-trivial branching process, and thus a non-trivial SOC state in BS model is 
   achieved only when the branching factor $r_b > 1$. To justify this idea we have considered a stochastic 
   version of the BS model where other than minimal fitness site, only one neighboring site is updated. This 
   model is numerically studied on $1d$ and $2d$ regular lattices, Barab\'asi-Albert scale-free networks and 
   $N$-clique graphs and in all these cases $r_b > 1$ and non-trivial SOC states are observed. In addition we 
   have have seen that in $1d$ where on the average 1.5 neighbors are updated in a mutation event, one still 
   has a SOC state. These evidences led us to conjecture that in a stochastic BS model studied on any arbitrary 
   graph where the average branching factor is greater than unity would lead to a non-trivial SOC state.

E-mail: manna@bose.res.in

\end {document}